\documentclass[10pt]{article}
\usepackage{euscript,amssymb,epsfig}
\usepackage{amsfonts,latexsym}
\usepackage{euscript,amssymb}
\usepackage{epsfig}

\def\RR{{\mathbb R}}

\catcode`\@=11
\def\lesssim{\mathrel{\mathpalette\vereq<}}
\def\vereq#1#2{\lower3pt\vbox{\baselineskip1.5pt \lineskip1.5pt
\ialign{$\m@th#1\hfill##\hfil$\crcr#2\crcr\sim\crcr}}}
\def\gtrsim{\mathrel{\mathpalette\vereq>}}

\def\Let@{\relax\iffalse{\fi\let\\=\cr\iffalse}\fi}
\def\vspace@{\def\vspace##1{\crcr\noalign{\vskip##1\relax}}}
\def\multilimits@{\bgroup\vspace@\Let@
 \baselineskip\fontdimen10 \scriptfont\tw@
 \advance\baselineskip\fontdimen12 \scriptfont\tw@
 \lineskip\thr@@\fontdimen8 \scriptfont\thr@@
 \lineskiplimit\lineskip
 \vbox\bgroup\ialign\bgroup\hfil$\m@th\scriptstyle{##}$\hfil\crcr}
\def\Sb{_\multilimits@}
\def\endSb{\crcr\egroup\egroup\egroup}
\def\Sp{^\multilimits@}

\newcommand{\be}[1]{\begin{equation}\label{#1}}
\newcommand{\ee}{\end{equation}}
\newcommand{\ba}[1]{\begin{eqnarray}\label{#1}}
\newcommand{\ea}{\end{eqnarray}}
\newcommand{\rf}[1]{(\ref{#1})}
\newcommand{\nn}{\nonumber}
\newcommand{\bmatrix}[1]{\left( \begin{array}{#1}}
\newcommand{\ematrix}{\end{array}\right)}

\newlength{\indentedwidth}
\newdimen\mathindent
\indentedwidth=\mathindent

\newenvironment{indented}{\begin{indented}}{\end{indented}}
\newenvironment{varindent}[1]{\begin{varindent}{#1}}{\end{varindent}}
\def\indented{\list{}{\itemsep=0\p@\labelsep=0\p@\itemindent=0\p@
   \labelwidth=0\p@\leftmargin=\mathindent\topsep=0\p@\partopsep=0\p@
   \parsep=0\p@\listparindent=15\p@}\footnotesize\rm}

\def\varindent#1{\setlength{\varind}{#1}%
   \list{}{\itemsep=0\p@\labelsep=0\p@\itemindent=0\p@
   \labelwidth=0\p@\leftmargin=\varind\topsep=0\p@\partopsep=0\p@
   \parsep=0\p@\listparindent=15\p@}\footnotesize\rm}

\def\etal{{\it et al\/}}

\begin{document}
\author{U. G\"unther$^a$\footnote{e-mail:
u.guenther@fz-rossendorf.de}~\footnote{present address: Research
Center Rossendorf, P.O. Box 510119, D-01314 Dresden, Germany}\, ,
A. Starobinsky$^b$\footnote{e-mail: alstar@landau.ac.ru}~
 \, and A. Zhuk$^c$\footnote{e-mail: zhuk@paco.net}\\[2ex] $^a$
Gravitationsprojekt, Mathematische Physik I,\\ Institut f\"ur
Mathematik, Universit\"at Potsdam,\\ Am Neuen Palais 10, PF
601553, D-14415 Potsdam, Germany
\\[1ex] $^b$ Landau Institute for Theoretical Physics, \\
Kosygina 2, Moscow 119334, Russia
\\[1ex] $^c$ Department of Physics, University of Odessa,\\ 2
Dvoryanskaya St., Odessa 65100, Ukraine }
\title{Multidimensional cosmological models: cosmological and
astrophysical implications and constraints}

\date{28.10.2003}
%%%%%%%%%%%%%%%%%%%%%%%%%%%%%%%%%
%
%\begin{document}
%
\maketitle
\begin{abstract} We investigate four-dimensional effective theories
which are obtained by dimensional reduction of multidimensional
cosmological models with factorizable geometry and consider the
interaction between conformal excitations of the internal space
(geometrical moduli excitations) and Abelian gauge fields. It is
assumed that the internal space background can be stabilized by
minima of an effective potential. The conformal excitations over
such a background have the form of massive scalar fields
(gravitational excitons) propagating in the external spacetime. We
discuss cosmological and astrophysical implications of the
interaction between gravexcitons and four-dimensional photons as
well as constraints arising on multidimensional models of the type
considered in our paper. In particular, we show that due to the
experimental bounds on the variation of the fine structure
constant, gravexcitons should decay before nucleosynthesis starts.
For a successful nucleosynthesis the masses of the decaying
gravexcitons should be $m \gtrsim 10^4$GeV. Furthermore, we
discuss the possible contribution of gravexcitons to UHECR. It is
shown that, at energies $E \sim 10^{20}$eV, the decay length of
gravexcitons with masses $m \gtrsim 10^4$GeV is very small, but
that for $m \lesssim 10^2$GeV it becomes much larger than the
Greisen-Zatsepin-Kuzmin cut-off distance. Finally, we investigate
the possibility for gravexciton-photon oscillations in strong
magnetic fields of astrophysical objects. The corresponding
estimates indicate
 that even the high magnetic field strengths $B$ of  magnetars (special
types of pulsars with $B>B_{critical}\sim 4.4 \times
10^{13}$Gauss) are not sufficient for an efficient and copious
production of gravexcitons.
\end{abstract}

\bigskip

\hspace*{0.950cm} PACS number(s): 04.50.+h, 11.25.Mj, 98.80.Jk

%\maketitle
%%%%%%%%%%%%%%%%%%%%%%%%%%%%%%%%%%%%%%%%%%%%%%%%%%%%%%%%%%%%%%%%%
%
%\newpage

\section{Introduction \label{intro}}

%\setcounter{equation}{0}
%%%%%
Multidimensionality of our Universe is one of the most intriguing
assumption in modern physics. It is a natural ingredient of
theories which unify different fundamental interactions with
gravity, such like string/M-theory \cite{pol-wit}. The idea has
received a great deal of renewed attention over the last few years
within the "brane-world" description of the Universe. In this
approach the $SU(3)\times SU(2)\times U(1)$ standard model (SM)
fields, related to usual four-dimensional physics, are localized
on a three-dimensional space-like hypersurface (brane) whereas the
gravitational field propagates in the whole (bulk) spacetime. The
geometry can be factorizable, as in the standard Kaluza-Klein (KK)
approach, or non-factorizable as in M-theory inspired
Randall-Sundrum (RS) scenarios \cite{RS}. For factorizable KK
geometries the topology is the product
\begin{equation}
\label{1.01}{  M}={  M}_0\times {  M}_1\times \dots \times { M}_n
\end{equation}
 of a non-warped
manifold ${  M}_0$, which constitutes the external spacetime, and
compact manifolds ${  M}_i ,\ i=1,\ldots,n$ as internal spaces
which are warped with functions depending on the external
coordinates. Factorizable models of this type will be the subject
of the present consideration\footnote{In contrast, the
non-factorizable geometries of RS-type models consist of two or
more patches of 5D anti-de Sitter space glued together along
branes (with one of them identified as our "world-brane"). The
$4-$dimensional spacetimes are then warped with factors which
depend on the extra $D^{\prime} =1$ dimension.}. For these models,
the four-dimensional Planck scale $M_{Pl(4)}$ and the fundamental
mass scale $M_{*(4+D^{\prime})}$ are connected by the relation
%%%%%
\be{1.2} M_{Pl(4)}^2 \sim V_{D^{\prime}}
M_{*(4+D^{\prime})}^{2+D^{\prime}}\, , \ee
%%%%%
where $V_{D^{\prime}}=\prod_{i=1}^n \mbox{\rm Vol} ({  M}_i)$
denotes the volume of the compactified $D^{\prime}=\sum_{i=1}^n
d_i $ extra dimensions $[d_i=\dim({  M}_i)]$. Relation \rf{1.2} is
valid for models with factorizable geometry regardless of SM
matter localized on branes, i.e. it is valid in
 KK scenarios as well as in brane-world scenarios. It reflects the
fact that gravitons propagate in the whole multidimensional bulk
${  M}$. The gravity force law in the product manifold ${  M}$
depends strongly on the considered length scale. Introducing the
characteristic sizes $b_i$  of the internal factor spaces as
$b_i\sim\mbox{\rm Vol} (M_i)^{1/d_i}$ and denoting the largest
size among the $b_i$ as $b_l=\max b_i $ and the smallest one by
$b_s=\min b_i$, it changes from $r^{-2}$ at scales $r> b_l$ in an
effectively $(d_0=4)-$dimensional spacetime ${  M}_0$  to
$r^{-(2+D')}$ at scales $r< b_s$ where all $D=d_0+D^{\prime}$
dimensions of the complete manifold ${  M}$ contribute to the
dilution of the effective gravity force (see e.g. Refs.
\cite{sub-mill1,sub-mill1b}). Physically acceptable values for the
compactification scales $b_i$ range from the four-dimensional
quantum gravity scale $M_{Pl(4)}^{-1}<b_s$ up to the recently
established experimental bound $b_l<10^{-1}$cm (see Ref.
\cite{experiment1,experiment2}).

For models with internal spaces all of the same characteristic
size $b$, and the fundamental energy scale set at
$M_{*(4+D^{\prime})} \sim$1TeV, this leads to ADD brane-world
scenarios \cite{sub-mill1} with compactification scales defined
from \rf{1.2} as
%%%%
\be{1.1} b \sim V_{D^{\prime}}^{1/D^{\prime}} \sim
10^{\frac{32}{D^{\prime}}-17} \mbox{cm}\, . \ee
%%%%
In these scenarios, physically acceptable values are constraint by
experiments  \cite{experiment1,experiment2} to $D^{\prime}\ge 3$
so that, e.g., for $D^{\prime} =3$ one arrives at a sub-micrometer
compactification scale $b\sim 10^{-5} \mbox{mm}$ of the internal
space. The fixing of $M_{*(4+D^{\prime})}$ at the scale $1-30$TeV
near the electroweak (EW) scale $M_{EW}\sim 246$ GeV \cite{ew}
provides an elegant resolution of the hierarchy problem. At the
same time, a scale hierarchy between $b$ and $
1\mbox{TeV}^{-1}\sim 10^{-17}$cm remains which only disappears in
the limit $D' \to \infty$ when $b \to
 1\mbox{TeV}^{-1}\sim 10^{-17}$cm.

%From other hand, in Kaluza-Klein approach the Standard Model
%fields, similar to the gravity, can propagate in whole
%multidimensional bulk. Such assumption results in the following
%experimental bounds for $b$: $b \leq l_{EW}$ . Thus, to solve the
%hierarchy between $b$ and $l_{EW}$, we can fix $b$ at the Fermi
%scales: $b \sim l_{EW}$. Then, from eq. \rf{1.2} we obtain an
%estimate for $(D=4+D')$--dimensional fundamental scale:
%%%%
%\be{0.3} M_{*(4+D^{\prime})} \sim 10^{(38+3D')/(2+D')}\mbox{GeV}\,
%.\ee
%%%%%
%For example,
%%%%%
%\be{0.4} M_{*(4+D^{\prime})} \vphantom{\int} \sim
%\left\{\begin{array}{rcl} 10^{11}\mbox{GeV}&{}&, \quad D'=2\\
%10^{7}\mbox{GeV}&{}&, \quad D'=6\\ 20 \mbox{TeV}&{}&, \quad
%D'=22\\
%\end{array}\right.
%\ee
%%%%%
%As can be easily seen from these estimates, the $D$--dimensional
%fundamental scale $M_{*(4+D^{\prime})}$ is close to the
%electroweak scale $M_{EW}\sim 1TeV$ if the number of the extra
%dimensions is sufficiently large. Thus, the hierarchy problem can
%be also solved in the Kaluza-Klein approach in the case of large
%number of additional dimensions (both scenarios, KK as well as
%ADD, approach to each other with the growth of number of the extra
%dimensions).

 In contrast to the 4D gravity force law, which is
experimentally tested at length scales above $0.1$ mm,  the
four-dimensionality of the SM gauge interactions is tested down to
scales of about $M_{EW}^{-1} \sim  l_{EW}\sim  10^{-16}$ cm. For
models with large extra dimensions, this requires a localization
of the SM fields on a world-brane. In general, this can be a
higher-dimensional $p-$brane with $p\ge 3$. The $d_{||}=p-3$
longitudinal dimensions
 of this $p-$brane  should then be compactified at sufficiently
small scales $b<l_{EW}$  \cite{ant1}.

According to observational data the extra-dimensional space
components  should be static or nearly static at least from the
time of primordial nucleosynthesis. Otherwise the fundamental
physical constants would vary (see e.g. \cite{GZ(CQG2),CV}). Eq.
\rf{1.2} shows, for example, that if $V_{D'}$ is a dynamical
function\footnote{It is clear that a dynamical behavior of the
volume of the extra dimensions allows for several different
implications of Eq. \rf{1.2}: (1) one can assume
$M_{*(4+D^{\prime})}\sim M_{EW}$ as the fundamental relation and
keeps it fixed for varying $M_{Pl(4)}$ \cite{sub-mill3}, (2) the
four-dimensional Planck scale $M_{Pl(4)}$ is the fundamental scale
and the multidimensional/electroweak scale
$M_{*(4+D^{\prime})}\sim M_{EW}$ is varying when $V_{D^{\prime}}$
varies, (3) all three scales $M_{Pl(4)}$, $M_{*(4+D^{\prime})}$,
$M_{EW}$ are varying before nucleosynthesis and their present
values are defined by string dynamical processes, (4) for fixed
$M_{Pl(4)}$ and $M_{*(4+D^{\prime})}$ the relation
$M_{*(4+D^{\prime})}\sim M_{EW}$ is the result of an earlier
dynamics and not a fundamental one.} which varies with time then
the effective four-dimensional gravitational constant will vary as
well.
 This means that, at the present evolutionary stage of the
Universe,  the compactification scale of the internal space should
either be stabilized and trapped at the minimum of some effective
potential, or it should vary sufficiently slowly (similar to the
slowly varying cosmological constant in the quintessence scenario
\cite{WCOS} (see also the reviews \cite{SahStar}, \cite{Peebles}))
so that the variations of derived parameters, like the variation
of the fine-structure constant $\alpha$, would meet their
observational bounds. In both cases, small fluctuations over
stabilized or slowly varying compactification scales (conformal
scales/geometrical moduli) are still possible.

The stabilization of extra dimensions (moduli stabilization) in
models with large extra dimensions (ADD models) was the subject of
numerous investigations (see e.g., Refs.
\cite{sub-mill3,sub-mill2,d2,PS,BEG})\footnote{In most of these
papers, the moduli stabilization was  considered without regard to
the energy-momentum localized on the brane. A brane matter
contribution was taken into account, e.g., in \cite{PS}.}. In the
corresponding  considerations, the product topology of a
$(4+D^{\prime })-$dimensional bulk spacetime was constructed from
Einstein spaces with scale (warp) factors depending only on the
coordinates of the external four-dimensional factor space. As a
consequence, the conformal excitations of the extra-dimensional
space components have the form of massive scalar fields living in
the external (our) spacetime. Within the framework of
multidimensional cosmological models (MCM) such excitations were
investigated in
\cite{GZ1,GZ,GZ(CQG1998),GZ(PRD2),GMZ,GMZ2}\footnote{See also Ref.
\cite{AD1}, where a decoupling of scale factor excitations and
inflaton was observed for a special solution subset of the
Einstein equations.}. In Ref. \cite{GZ1} they were called
gravitational excitons. Later, since the ADD compactification
approach these geometrical moduli excitations are known as radions
\cite{sub-mill3,sub-mill2}.

In the present paper we study the interaction of gravitational
excitons with Abelian gauge fields, and in particular with
electro-magnetic (EM) fields. A possible observation of reactions
in this interaction channel would be of great interest because it
could provide strong evidence for the existence of extra
dimensions. The corresponding interaction term of the
four-dimensional effective theory has the form
\be{1.3} \Delta S_{EM}\sim \kappa_0 \int_{{  M}_0}d^4 x \sqrt{|\tilde{g}^{(0)}|}\psi F_{\mu
\nu} F^{\mu \nu},
\ee
where $F$ denotes the EM field strength tensor, and the massive
scalar field $\kappa_0\psi < 1$ describes small scale factor (warp
factor) excitations (gravexcitons) of the extra-dimensional space
components. The interaction term \rf{1.3} follows from a KK-like
dimensional reduction scheme, where
 the scale factors, their excitations $\psi$ and the EM
 field strength $F$ are considered in a
zero-mode approximation. As result they will only depend on the
coordinates of the external spacetime ${  M}_0$. In this scheme
the extra-dimensional space components should be compactified at
scales $b_i<l_{EW}$. (In a brane-world context one could interpret
such a setup as rough approximation of a $p-$brane endowed with an
internal warped product structure where three dimensions and the
time are represented by ${  M}_0$, the remaining $p-3$
longitudinal dimensions are mimicked by extra dimensional warped
factor spaces ${  M}_1\times \dots \times {  M}_n$, and where any
gravity contributions of the large transverse extra dimensions are
neglected.) Furthermore, it is assumed that the scale factors are
stabilized and frozen in one of the minima of an effective
potential with $\psi$ as fluctuations over this minimum. The main
goal of our paper consists in the investigation of interaction
\rf{1.3} and its cosmological and astrophysical implications.

The paper is structured as follows. In section \ref{excitons}, we
explain the general setup of our model and give a basic
description of gravitational excitons from extra dimensions. A
consideration of the interaction between gravitational excitons
and four-dimensional photons is presented in section \ref{gauge}.
It is followed by a discussion of cosmological and astrophysical
implications of this interaction (section \ref{astro}). Due to the
Planck scale suppression $\kappa_0 \sim 1/M_{Pl(4)}$ and a decay
rate $\Gamma \sim m^3/M_{Pl(4)}^2$, gravitational excitons with
mass $m$ are WIMPs (Weakly-Interacting Massive Particles
\cite{KT}) similar to
 other moduli fields, Polonyi fields
\cite{CFKRR,ENQ,CCQR,BKN,BBS} and scalarons \cite{Star1980}. We
investigate gravexcitons from a cosmological perspective by taking
into account experimental bounds on the variation of the fine
structure constant $\alpha$. Additionally, we discuss in this
section a possible gravexciton contribution to UHECR as well as
possible gravexciton-photon oscillations in strong magnetic fields
of magnetars. The main results are summarized in the Conclusion
 section.

%%%%%%%%%%%%%%%%%%%%%%%%%%%%%%%%%%%%%%%%%%%%%%%%%%%%%%%%%%%%%%%%%%%%

\section{Gravitational excitons \label{excitons}}

In this section we present a sketchy outline of the basics of
gravitational excitons from extra dimensions. A more detailed
description can be found, e.g.,  in our paper \cite{GZ1}.

Let us consider a multidimensional spacetime manifold ${  M}$ with
warped product topology \rf{1.01} and metric
\begin{equation}
\label{2.2}g=g_{MN}(X)dX^M\otimes
dX^N=g^{(0)}+\sum_{i=1}^ne^{2\beta ^i(x)}g^{(i)},
\end{equation}
where $x$ are some coordinates of the $(D_0=4)-$dimensional  manifold $%
{  M}_0 $ and
\begin{equation}
\label{2.3}g^{(0)}=g_{\mu \nu }^{(0)}(x)dx^\mu \otimes dx^\nu .
\end{equation}
%%%%%
Let further the internal factor manifolds ${  M}_i$ be
$d_i$-dimensional warped Einstein spaces with  warp factors
$e^{\beta^i (x)}$ and metrics $g^{(i)}=g^{(i)}_{m_in_i}(y_i)\,
dy_i^{m_i}\otimes dy_i^{n_i}$, i.e.,
%%%%%%
\begin{equation}
\label{2.4}R_{m_in_i}\left[ g^{(i)}\right] =\lambda
^ig_{m_in_i}^{(i)},\qquad m_i,n_i=1,\ldots ,d_i
\end{equation}
and
\begin{equation}
\label{2.5}R\left[ g^{(i)}\right] =\lambda ^id_i\equiv R_i \sim k
b_i^{-2}
\end{equation}
with $k = 0,\pm 1$. $b_i$
%= \left( \int d^{d_i}y\sqrt{|g^{(i)}|}\right)^{1/d_i}$
is the characteristic size of the Einstein space ${  M}_i$ (modulo
the dimensionless warp factor $a_i \equiv \exp ( \beta^i)$), i.e.
$b_i$ is the effective scale factor of the space ${  M}_i$ with
metric $g^{(i)}$ and corresponding volume\footnote{The volume is
well defined for positive curvature spaces ($k=+1$). For compact
negative and zero curvature spaces ($k=-1,0$), i.e. compact
hyperbolic spaces (CHSs) $M_i=H^{d_i}/\Gamma_i$ and tori
$T_j=R^{d_j}/\Gamma_j$, we interpret this volume as scaled volume
of the corresponding fundamental domain ("elementary cell")
$V_{d_i} \sim b_i^{d_i}\times V_{FD(i)}$ (see, e.g., \cite{SS} and
references therein).  Here $H^{d_i}$, $R^{d_j}$ are hyperbolic and
flat universal covering spaces, and $\Gamma_i$, $\Gamma_j$ ---
appropriate discrete groups of isometries. Furthermore, we assume
for the scale factors of the metrics $\gamma^i\sim b_i
\hat{\gamma}^i$, with $\hat{\gamma}^i$ scaled in such a way that
$V_{FD(i)}\sim \mathcal{O}(1)$. Thus, the volume $V_{d_i}$ is
mainly defined by $b_i$.}
%%%%%
\be{2.5b} V_{d_i } \equiv \int\limits_{M_i}d^{d_i}y
\sqrt{|g^{(i)}|} \sim b_i^{d_i}\quad i=1, \dots , n \, , \ee
%%%%%
where $V_{d_i}$ has dimensions $\mbox{cm}^{d_i}$.  Without loss of
generality we set the compactification scales of the internal
spaces at present time at $\beta^i = 0 \longrightarrow a_i = 1 \;
(i = 1,\ldots ,n)$. This means that at present time the total
volume of the internal spaces is completely defined by the
characteristic scale factors $b_i$:
%%%%%
\be{2.5c} V_{D^{\prime }} \equiv
\prod_{i=1}^n\int\limits_{M_i}d^{d_i}y \sqrt{|g^{(i)}|} \sim
\prod_{i=1}^n b_i^{d_i}\; . \ee
%%%%%
%where $V_{D^{\prime }}$ has dimensions $\mbox{cm}^{D'}$.
%$\mathcal{O}(m^{-d_i})$.

With total dimension $D=D_0+D'=D_0+\sum_{i=1}^nd_i$, \ $\kappa_D^2
= 8\pi / M^{2+D^{\prime}}_{*(4+D^{\prime})}$ \ --- \ a
$D$-dimensional gravitational constant, $\Lambda $ \ --- \ a
$D$-dimensional bare cosmological constant and $S_{YGH}$ the
standard York-Gibbons-Hawking boundary term, we consider an action
of the form
\begin{equation}
\label{2.6}S=\frac
1{2\kappa_D^2}\int\limits_Md^DX\sqrt{|g|}\left\{ R[g]-2\Lambda
\right\} +S_{m}+S_{YGH}.
\end{equation}
Here, the action $S_{m}$ corresponds to  possible matter fields,
e.g. gauge fields and scalar fields\footnote{Here, the action
$S_{m}$ is  treated as action component of matter fields in a
standard KK scenario. In an ADD-like large-scale compactification
scenario, such an ansatz can be used in the case when the
energy-momentum localized on the brane can be ignored in
comparison with the energy-momentum of the matter in the bulk
(see, e.g., Ref. \cite{GZ(PRD2),GMZ,GMZ2}). Then $S_m$ can provide
an approximate description of the bulk matter components.}. In
some models these matter fields can be considered
phenomenologically as a perfect fluid with energy density $\rho$
and corresponding equation of state \cite{GZ,GZ(CQG1998),GIM,KZ}.
This provides, for example, an efficient way to take into account
the Casimir effect \cite{CS}, Freund-Rubin monopoles
\cite{GMZ2,FR,AGHK} or other hypothetical potentials \cite{GZ,BZ}.
All these cases can be described  by an additional potential term
\begin{equation}
\label{2.7}S_{m}=-\int\limits_Md^DX\sqrt{|g|}\rho (x)\, .
\end{equation}
For some specific classes of models, e.g. for such describing the
Casimir effect or  Freund-Rubin monopoles, the energy density
$\rho $ depends on the external coordinates only through the scale
factors $a_i(x)=e^{\beta ^i(x)}\ (i=1,\ldots ,n)$ of the internal
factor spaces $M_i$. Obviously, the functional dependence on these
scale factors is highly model depending. Throughout the present
section we will not specify this dependence, using instead the
general expression \rf{2.7}. An explicit consideration of matter
fields is left for sections \ref{gauge} and \ref{astro}.

After dimensional reduction and conformal transformation
\begin{equation}
\label{2.10}g_{\mu \nu }^{(0)}=\Omega ^2\tilde g_{\mu \nu }^{(0)},
\end{equation}
\begin{equation}
\label{2.11}\Omega =\exp\left( -\frac 1{D_0-2}\sum_{i=1}^nd_i\beta
^i\right)\,
\end{equation}
from the intermediate Brans-Dicke frame to the final Einstein
frame,  action \rf{2.6} reads \cite{GZ1,16}
%%%%%
\begin{equation}
\label{2.12}S[\tilde g^{(0)},\beta ] = \frac 1{2\kappa
_0^2}\int\limits_{M_0}d^{D_0}x\sqrt{|\tilde g^{(0)}|}\left\{
\tilde R\left[ \tilde g^{(0)}\right] -\bar G_{ij}\tilde g^{(0)\mu
\nu }\partial _\mu \beta ^i\,\partial _\nu \beta
^j-2U_{eff}\right\} \, ,
\end{equation}
%%%%%
where
%%%%
\be{2.8a}\kappa _0^2=\kappa_D ^2/V_{D'}= 8\pi/M^2_{Pl}\quad
\Longrightarrow \quad M_{Pl}^2 = V_{D^{\prime}}
M_{*(4+D^{\prime})}^{(2+D^{\prime})}\ee
%%%%
is the $D_0$-dimensional gravitational constant (hereafter,
$M_{Pl} \equiv M_{Pl(4)}$), and formula \rf{2.8a} reproduces
equation \rf{1.2}. The tensor components of the midisuperspace
metric (target space metric on $\RR _T^n$) $\bar G_{ij}\
(i,j=1,\ldots ,n)$, its inverse metric $\bar G^{ij} $ and the
effective potential are given as
\begin{equation}
\label{2.13}\bar G_{ij}=d_i\delta _{ij}+\frac 1{D_0-2}d_id_j\, ,
\quad \bar G^{ij}=\frac{\delta ^{ij}}{d_i}+\frac 1{2-D}
\end{equation}
and
\begin{equation}
\label{2.15}U_{eff} ( \beta ) ={\left( \prod_{i=1}^ne^{d_i\beta
^i}\right) }^{-\frac 2{D_0-2}}\left[ -\frac
12\sum_{i=1}^nR_ie^{-2\beta ^i}+\Lambda +\kappa_D ^2\rho \right]
\, .
\end{equation}

We recall that $\rho $ depends on the scale factors of the
internal spaces: $\rho =\rho \left( \beta ^1,\ldots ,\beta
^n\right) $. Thus, action \rf{2.12} describes a self-gravitating
$\sigma -$model with flat target space $(\RR _T^n,\bar G)$
(\ref{2.13}) and self-interaction potential (\ref{2.15}).
Accordingly, the internal spaces  can stabilize if the effective
potential (\ref{2.15}) has at least one minimum with respect to
the scale factors $\beta^i$. Because the conformal transformation
(\ref{2.10}) was performed only with respect to the external
metric $g^{(0)}$, the stability of the internal space
configurations does not depend on the concrete choice of the frame
(Einstein or Brans-Dicke).

In the following we consider models with a constant scale factor
background which is localized in a minimum of the effective
potential $U_{eff}$. The values of the scale factors in this
minimum are rescaled in such a way that at present time it holds:
$\vec \beta =0,\quad \left. \frac{\partial U_{eff}}{\partial \beta
^i}\right| _{\vec \beta =0}=0$.  In general, the effective
potential $U_{eff}$ can have more than one minimum so that
transitions between these minima should be possible. For
simplicity, we leave effects related to such transitions out of
the scope of our present investigation. Instead, we will
concentrate on small scale factor fluctuations $\beta^i < 1$ in
the vicinity of one minimum only. The action functional \rf{2.12}
can then be rewritten in terms of decoupled normal modes
$\kappa_0\psi^i < 1$ (for details we refer to
\cite{GZ1,GZ,GZ(CQG1998)}):
%%%%
\begin{eqnarray}\label{2.30}
S[\tilde g^{(0)},\psi ] & = & \frac{1}{2\kappa _0^2}\int
\limits_{M_0}d^{D_0}x \sqrt {|\tilde g^{(0)}|}\left\{\tilde
R\left[\tilde g^{(0)}\right] - 2\Lambda _{eff}\right\} +
\nonumber\\ \ & + & \sum_{i=1}^{n}\frac{1}{2}\int
\limits_{M_0}d^{D_0}x \sqrt {|\tilde g^{(0)}|}\left\{-\tilde
g^{(0)\mu \nu}\psi ^i_{,\mu}\psi ^i_{,\nu} - m_{i}^2\psi ^i\psi
^i\right\}\, ,
\end{eqnarray}
%%%%%
where $\Lambda _{eff}\equiv U_{eff}(\vec \beta =0) $ plays the
role of a $D_0$-dimensional effective cosmological constant. The
normal modes and their masses squared $m_i^2$ are obtained by a
simultaneous diagonalization of the $\sigma -$model metric
\rf{2.13} and the Hessian
%%%%
\begin{equation}
\label{2.26}\left. \frac{\partial ^2U_{eff}}{\partial \beta
^i\,\partial \beta ^k}\right| _{\vec \beta =0}\; .
\end{equation}
%%%%
In the special case of only one internal space ($n=1$), this
procedure reduces to a simple rescaling
%%%%%
\be{2a}
\beta^1 = - \kappa_0 \sqrt{\frac{D_0-2}{d_1(D-2)}}\psi^1 \ ,
\ee
%%%%%
%where, later on, we shall use for definiteness sign minus,
and
%%%%
\be{2aa}
m_1^2 = \frac{D_0-2}{d_1(D-2)}\left. \frac{\partial^2
U_{eff}}{\partial {\left(\beta^1\right)}^2}\right|_{\beta^1 =0}\,
.
\ee
%%%%

{}Summarizing this section, we conclude that conformal zero-mode
excitations of the internal factor spaces $M_i$ have the form of
massive scalar fields developing on the background of the external
spacetime $M_0$. In analogy with excitons in solid state physics
(excitations of the electronic subsystem of a crystal), we called
these conformal excitations of the internal spaces gravitational
excitons \cite{GZ1}. Later, since Refs. \cite{sub-mill3,sub-mill2}
these particles are also known as radions.

%\vspace{0.5cm}
%%%%%%%%%%%%%%%%%%%%%%%%%%%%%%%%%%%%%%%%%%%%%%%%%%%%%%%%%%%%%%%%%%%%

\section{Abelian gauge fields\label{gauge}}

In this section we study the interaction of gravitational excitons
with  Abelian gauge fields, and in particular with the standard
electromagnetic field of $U(1)_{EM}$ symmetry. Strictly speaking,
the photon will not exist as a separate gauge boson at
temperatures higher than the electroweak scale $M_{EW}\sim 246$
GeV where the full electroweak  $SU(2)\times U(1)$ model should be
considered. Nevertheless, our results should reproduce the correct
coupling term between the gravexciton sector and the EM gauge
field sector of the theory. In the next section we will use this
coupling term for estimating the strength of cosmological and
astrophysical effects related to the corresponding interaction
channel.

In order to derive the concrete form of the coupling term in the
dimensionally reduced, four-dimensional effective theory we start
from the simplified toy model ansatz
\be{1e0}
S_{EM} =  -\frac12 \int_M d^DX\sqrt {|g|}F_{MN}F^{MN} \; ,
\ee
where the gauge field is assumed Abelian also in the
higher-dimensional setup. Additionally, we work in the zero mode
approximation for these fields, i.e. we keep only the zero modes
of the harmonic expansion in mass eigenstates of the
higher-dimensional fields\footnote{The excitation of Kaluza-Klein
modes of Abelian gauge fields was considered, e.g., in Ref.
\cite{MUM}.}, \cite{18,21}. In this case the Abelian vector
potential depends only on the external coordinates, $A_M=A_M(x),\
\ (M=1,\ldots ,D)$, and the corresponding non-zero components of
the field strength tensor are: $F_{\mu \nu }=\partial _\mu A_\nu
-\partial _\nu A_\mu, \ \ (\mu ,\nu =1,\ldots ,D_0)$, and $F_{\mu
m_i}=\partial _\mu A_{m_i}-\partial _{m_i}A_\mu =\partial _\mu
A_{m_i},\ \ (m_i=1,\ldots ,d_i;i=1,\ldots ,n)$.

Dimensional reduction of the gauge field action \rf{1e0} yields
\be{1e}
S_{EM}  =  -\frac12 \int_{M_0} d^{D_0}x\sqrt {|g^{(0)}|}
\prod_{i=1}^n e^{d_i\beta ^i} \left\{ F_{\mu \nu }F^{\mu \nu}
+2g^{(0)\mu \nu } \sum_{i=1}^n e^{-2\beta^i (x)} \bar
g^{(i)m_in_i}
\partial_{\mu } A_{m_i}\partial_{\nu } A_{n_i} \right\} \, ,
\ee
where we introduced the metric integral
\be{2e}\bar g^{(i)m_in_i}:= \frac 1{V_{d_i}}\int_{M_i}d^{d_i}y\sqrt{%
|g^{(i)}|}g^{(i)m_in_i}(y^i)
\ee
and included the factor $\sqrt{V_{D'}}$ into $A_M$ for
convenience: $\sqrt{V_{D'}} A_M\rightarrow A_M$. Due to this
redefinition, the field strength tensor $F_{\mu \nu}$ acquires the
usual dimensionality $\mbox{cm}^{-D_0/2}$ (in geometrical units
$\hbar = c =1$). In Eq. \rf{1e} we assumed $F^{\mu \nu }=g^{(0)\mu
\kappa }g^{(0)\nu \delta }F_{\kappa \delta }$.

It is easily seen that the $A_{m_i}$ components play the role of
scalar fields in the $D_0$-dimensional spacetime. In what follows,
we will not investigate the dynamics of these fields. Instead, we
will concentrate on the interaction between gravexcitons and the
2-form field strength $F=dA,\ A=A_\mu dx^\mu $ which is described
by the first term of the action functional \rf{1e}. The
corresponding truncated action (without $A_{m_i}$ terms) will be
denoted by $\bar {S}_{EM} $.

The exact field strength 2-form $F=dA$ with components $F_{\mu \nu
}$ is invariant under gauge transformations $A\mapsto  A^f=A+df$ ,
$F^f=dA+d^2f=dA=F$, with $f(x)$ any smooth and single-valued
function. Accordingly, $\bar {S}_{EM} $ is gauge invariant too
[see \rf{1e}].

The action functional \rf{1e} is written in Brans-Dicke frame. For
passing by the conformal transformation \rf{2.10}, \rf{2.11} to
the Einstein frame we choose an ansatz
\begin{equation}
\label{3e}A =\Omega ^k\tilde A \,
\end{equation}
for the vector potential and introduce the auxiliary field
strength $\bar F$ by the relation
\ba{3e1}F&=&dA=d(\Omega^k \tilde A)=\Omega^k \bar F\; ,\nn \\
\bar F&=& d(\ln \Omega^k )\wedge \tilde A + d \tilde A \; .
\ea
The conformally transformed effective action reads then
\begin{equation}\label{4e}
\bar{S}_{EM}=-\frac12 \int_{M_0} d^{D_0}x \sqrt{|\tilde g^{(0)}|}
\left\{ \Omega^{2(k-1)} \bar F_{\mu \nu }\bar F^{\mu \nu }
\right\} \, ,
\end{equation}
where the external space indices are raised and lowered by the
metric $\tilde g^{(0)}$. With  $\tilde F=d\tilde A$, we have in
\rf{4e} explicitly
\ba{7e}
\bar F_{\mu \nu }\bar F^{\mu \nu } &=& \tilde F_{\mu \nu }\tilde
F^{\mu \nu } - 2 \tilde F^{\mu \nu } \left[ \tilde A_{\mu }
\partial_{\nu }(\ln \Omega ^k) - \tilde A_{\nu } \partial_{\mu }
(\ln \Omega ^k) \right]  \\ & & + 2\left[ \tilde g^{(0)\mu \kappa
}\partial_{\mu } (\ln \Omega ^k)\partial_{\kappa }(\ln \Omega ^k)
\tilde A^{\nu }\tilde A_{\nu } - \left( \tilde A^{\mu
}\partial_{\mu }(\ln \Omega ^k)\right)^2 \right]  . \nn
\ea

In order to fix the conformal weight $k$ of the vector potential
in \rf{3e}, we require the effective external field strength
tensor $ \bar F_{\mu \nu } $ in \rf{4e} to be gauge invariant,
i.e. to be invariant under $\tilde A \mapsto \tilde A^f=\tilde
A+df$. {}From \rf{3e1} we have for this transformation
\ba{10e}
\bar F \mapsto \bar F^f & = & d\tilde A + d^2f + d(\ln \Omega
^k)\wedge (\tilde A\ +df) \\[1ex]
 & = & \bar F + d(\ln \Omega ^k)\wedge df \nn
\ea
so that for nontrivial $\Omega \neq 1$ the gauge invariance $\bar
F=\bar F^f$ is only achieved for zero conformal weight $k=0$. The
same result follows also directly from the gauge invariance  of
the field strength tensor $F$ in \rf{1e} and the ansatz \rf{3e}:
One checks immediately that $\bar F$ is invariant under a
transformation $\tilde A\mapsto \check A=\tilde A+\Omega ^{-k}df,$
which only for $k=0$ is a gauge transformation.

This means that in order to preserve the gauge invariance of the
action functional, when passing from the Brans-Dicke frame to the
Einstein frame, we have to keep the vector potential unchanged,
i.e. we have to fix the conformal weight at $k=0$. As result we
arrive at the action functional
%%%%%%
\be{11e} \bar{S}_{EM}= -\frac12
\int_{M_0} d^{D_0}x \sqrt{|\tilde g^{(0)}|} \left\{
e^{\frac{2}{D_0-2}\sum ^n_{i=1}d_i\beta ^i(x)} F_{\mu \nu } F^{\mu
\nu }  \right\}
\ee
%%%%%%
with  dilatonic coupling of the Abelian gauge fields to the
gravitational excitons.

For completeness we note that
 for $k=1$, according to \rf{4e} and \rf{7e}, we obtain
 a theory with a pure free action term
$\tilde F_{\mu \nu }\tilde F^{\mu \nu }$ without any prefactor
$(\Omega ^{2(k-1)}=1)$ but with explicitly destroyed gauge
invariance. The corresponding
 effective action reads
\ba{12e} \bar{S}_{EM}&=& -\frac12 \int_{M_0} d^{D_0}x \sqrt{|\tilde
g^{(0)}|} \left\{ \tilde F_{\mu \nu }\tilde F^{\mu \nu } - 4\tilde
F^{\mu \nu } \tilde A_{[\mu }
\partial_{\nu ]} \left( \ln \Omega \right) \right. \nonumber \\
&+& \left. 2\left[ \tilde g^{(0)\mu \kappa }\partial_{\mu } \left(
\ln \Omega \right) \partial_{\kappa } \left( \ln \Omega \right)
\tilde A^{\nu }\tilde A_{\nu } - \left( \tilde A^{\mu
}\partial_{\mu } \ln \Omega \right)^2 \right]
 \right\}  .
\ea

Obviously, the localization of the scale factors $\beta^i$ at
their present values $\beta^i =0$ results in $\Omega \equiv 1$.
Then, both approaches \rf{11e} and \rf{12e} coincide with each
other. However, the presence of small scale factor fluctuations
above this background will restore the dilatonic coupling of
\rf{11e} (see also the next section).

%%%%%%%%%%%%%%%%%%%%%%%%%%%%%%%%%%%%%%%%%%%%%%%%%%%%%%%%%%%%%%%%%
\section{Gravitational excitons and their cosmological and astrophysical
implications \label{astro}}
%\bigskip
%\setcounter{equation}{0}

\begin{figure}[htb]
\begin{center}
\epsfig{file=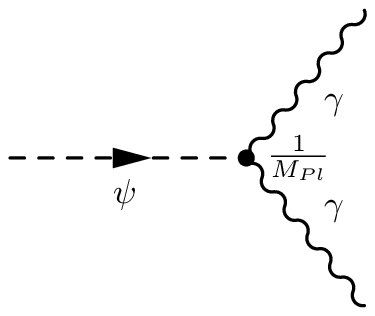,angle=0, width=4.5cm}
\end{center}
\caption{\label{fig1} Planck-scale suppressed gravexciton decay:
$\psi \longrightarrow 2 \gamma$.}
\end{figure}

In this section we discuss some cosmological and astrophysical
implications related to the possible existence of gravitational
excitons. We suppose that the scale factor background of the
internal spaces is localized in one of the minima of the effective
potential (see section 2) and that gravexcitons are present as
small fluctuations above this static background. Our analysis is
based on the dilatonic coupling \rf{11e} which describes the
interaction between gravexcitons and zero mode
photons\footnote{Brane-world models with on-brane dilatonic
coupling terms have been considered, e.g., in Refs.
\cite{boehm,IJMPD}. In a rough approximation, the results of the
present section will also hold for these models.}.
 Hereafter, we treat these KK zero mode photons as our
usual SM matter photons.
 In particular,
the vector potential ${A_{\mu}}(x)$ of the previous section
corresponds to our 4D photons.
%For this purpose
In the following we consider the simplest example --- the
interaction between gravitational excitons and photons in a system
with only one internal space $(n=1)$ with its scale factor
$\beta^1$  localized in one of the minima of the effective
potential \rf{2.15}. We rescale the size of the corresponding
factor space in such a way that the background component takes the
value $\beta^1 =0$ at present time. Then, for small scale factor
fluctuations $\beta^1 < 1$ action \rf{2.30} together with \rf{11e}
reads
\begin{eqnarray}\label{1a}
S & = & \frac{1}{2\kappa _0^2}\int \limits_{M_0}d^{D_0}x \sqrt
{|\tilde g^{(0)}|}\left\{\tilde R\left[\tilde g^{(0)}\right] -
2\Lambda _{eff}\right\} + \nonumber\\ \ & + & \frac{1}{2}\int
\limits_{M_0}d^{D_0}x \sqrt {|\tilde g^{(0)}|}\left\{-\tilde
g^{(0)\mu \nu}\psi _{,\mu}\psi _{,\nu} - m_{\psi}^2\psi \psi
\right\} - \nonumber\\ \ & - & \frac{1}{2}\int
\limits_{M_0}d^{D_0}x \sqrt {|\tilde g^{(0)}|}\left\{ F_{\mu
\nu}F^{\mu \nu} - 2\sqrt{\frac{d_1}{(D_0-2)(D-2)}}\kappa_0 \psi
F_{\mu \nu} F^{\mu \nu}\right\} + \ldots \ .
\end{eqnarray}
We used the notations of eq. \rf{2.30}, $m_{\psi} := m_{1}$ and
relation \rf{2a} between $\beta^1$ and the rescaled fluctuational
component  $\psi\equiv \psi^1$. As mentioned above, $\kappa^2_0 =
8\pi / M^2_{Pl} $ is the $D_0$-dimensional (usually $D_0 = 4$)
gravitational constant. The last term describes the interaction
between gravitational excitons and photons. In lowest order
tree-level approximation this term corresponds to the vertex of
Fig. \ref{fig1}

%\noindent
%\begin{picture}(16.5,4.5)
%\put(4.5,4.5){\special{em:graph figura01.pcx}}
%\end{picture}

%\vspace{3.5cm}

%\vspace*{-2.5cm}

\noindent and describes the decay of a gravitational exciton into
two photons. The probability of this decay is easily  estimated
as\footnote{Exact calculations give $\Gamma = \left[ \,
2d_1/(d_1+2)\right]\left(m_{\psi}^3/M_{Pl}^2\right)$.}
\be{3a}
\Gamma \sim \left(\frac{1}{M_{Pl}}\right)^2 m_{\psi}^3 =
\left(\frac{m_{\psi}}{M_{Pl}}\right)^3\frac{1}{t_{Pl}} \ll
m_{\psi}\ ,
\ee
which results in a life-time $\tau$ of the gravitational excitons
with respect to this decay channel of
\be{4a}
\tau = \frac{1}{\Gamma} \sim
\left(\frac{M_{Pl}}{m_{\psi}}\right)^3 t_{Pl} \, .
\ee
Similar to Polonyi particles in spontaneously broken supergravity
\cite{CFKRR,ENQ}, scalarons in the $(R +R^2)$ fourth order theory
of gravity \cite{Star1980} or moduli fields in the hidden sector
of SUSY \cite{CCQR,BKN,BBS}, gravitational excitons are WIMPs
(Weakly-Interacting Massive Particles \cite{KT}) because their
coupling to the observable matter is suppressed by powers of the
Planck scale.

\subsection{Light gravexcitons\label{light}}

We first consider light excitons with masses $m_{\psi} \le
10^{-21} M_{Pl} \sim 10^{-2}\mbox{GeV} \sim 20 m_e$ (where $m_e$
is the electron mass). Their life-time $\tau $ is greater than the
age of the universe $\tau \ge 10^{19}\mbox{sec}
> t_{univ} \sim 10^{18}$ sec. From this estimate follows that
 they can be considered as Dark Matter (DM). The type of the DM
depends on the masses of the particles which constitute it. It is
hot for $m_{DM} \le 50 - 100$ eV, warm for $ 100 \, \mbox{eV} \le
m_{DM} \le 10$ keV and cold for $m_{DM} \ge 10 - 50$ keV.
%In the approach where the
%$D-$dimensional gravitational constant is defined as $\kappa_D^2 \sim
%M_{Pl}^{-(2+D^{\prime})}$, the gravexciton masses are related to the
%compactification scales by $ m_{\psi} \sim
%\left( a_{(c)} \right)^{-(D-2)/(D_0-2)}$ (see Refs. \cite{1a,1b,GZ(PRD2)}).

Concerning the gravexciton decay processes we note that there
exists a characteristic time $t_D$ within the evolution of the
Universe when these processes are most efficient (see e.g. eq.
\rf{13a} below). The time can be estimated as: $H(t_D) \sim \Gamma
\, \Longrightarrow t_D \sim \tau$, where $H = \dot a / a $ is the
Hubble constant, $a$
--- the scale factor of the external space in the Einstein frame
and dots denote derivatives with respect to the synchronous time
$t$ in the Einstein frame. It is clear that for gravexcitons with
masses $m_{\psi} \ll 10^{-2}$GeV we can neglect their decay during
the evolution of the Universe
 because for these masses the characteristic time $t_D$ is much greater than the
age of our Universe. Such gravexcitons up to present time undergo
coherent oscillations according to the approximate equation
%%%%%%%
\be{5a} \ddot \psi + 3H\dot \psi +m_{\psi}^2\psi = 0\, , \ee
%%%%%%%
and  did not convert into radiation with subsequent reheating of
the Universe\footnote{Strictly speaking, the decay $\Gamma$ of
gravexcitons as well as their production due to interaction with
other matter (with high temperature radiation at the radiation
dominated stage, in particular) have to be taken into account
here, too. The latter process can be described by a source term in
the rhs of the effective equation of motion
$$ \ddot \psi + (3H +\Gamma) \dot \psi +m_{\psi}^2\psi = I_m ,$$
where $I_m$ is proportional either to the trace of the EMT of
matter on the world-brane (in brane-world scenarios \cite{PS}), or
to the energy density of matter in our 4D Universe (in the
Kaluza-Klein approach
\cite{GZ(CQG2),GZ(CQG1998),GIM,A.ZhukCQG13(1996)2163}). The
behavior of the source term $I_m$ is defined not only by the
red-shifting EMT of matter but also by gravexcitons' interactions
with it, e.g. by gravexciton decay into radiation as well as by
reactions of the type: gravexciton + $\gamma \to$ everything. The
thermal production and balance of gravexcitons is described by the
Boltzmann equation for their number density  $n_{\psi}$:
$$ \frac{dn_{\psi}}{dt} +3Hn_{\psi} = -\langle\sigma v\rangle n_{\gamma}
(n_{\psi}-n_{\psi(eq)}) -\Gamma (n_{\psi}-n_{\psi(eq)})\, ,$$
where $\langle\sigma v\rangle \sim M_{Pl}^{-2}$ is the typical
cross section of these reactions and $\Gamma$ is given by Eq.
\rf{3a}. $n_{\psi(eq)}$ denotes the gravexciton number density in
thermal equilibrium (The subscript is not to be mixed with that of
the characteristic time of matter/radiation equality). However,
due to extremely weak interaction of gravexcitons with matter, it
is very difficult for them to reach the thermal equilibrium. For
particles which never dominate the Universe, $\psi$ usually starts
to oscillate when $n_{\psi}<<n_{\psi(eq)}$ and, as it was shown in
\cite{kolb2}, the contribution of thermally produced gravexcitons
to the total energy density is negligible. Thus, in this case the
source term (that effectively reflects the efficiency of
interaction between gravexcitons and ordinary matter) does not
play any important role in Eq. \rf{5a}. On the other hand, in
scenarios where gravexcitons dominate the early dynamics of the
Universe their number density can initially be
$n_{\psi}>>n_{\psi(eq)}$ and it will decrease until it reaches its
thermal equilibrium at $n_{\psi(eq)}$. In such a scenario, the
initial amplitude can be assumed as $\psi_{in} \sim M_{Pl}$. Then
the number density of $\gamma$ quanta will be $n_{\gamma} \sim T^3
<< n_{\psi}$, so that $n_{\psi} \sim e^{-\Gamma t}a^{-3}$.
Therefore, thermal production of gravexcitons may be neglected.
The scenario when gravexcitons have enough time to decay into
radiation with subsequent reheating of the Universe is considered
below, in the subsection \ref{heavy}.}. Assuming that at the
present time the gravexciton energy density $\rho_{\psi}$ is less
or similar to the critical density $\rho_{\psi} \lesssim \rho_c$,
we obtain an upper bound for  the gravexciton masses
%\cite{GZ(PRD2)}
%%%%%%
\be{6a} m_{\psi} \lesssim \sqrt{\frac{3}{8\pi}} H_{eq}
\left(\frac{M_{Pl}}{\psi_{in}} \right)^4 \sim 10^{-56} M_{Pl}
\left(\frac{M_{Pl}}{\psi_{in}} \right)^4\, , \ee
%%%%%
if $m_{\psi} > 10^{-56}M_{Pl}$, so that the oscillations began at
the RD stage\footnote{\label{Heq}In eq. \rf{6a} $H_{eq}$ denotes
the Hubble parameter at that evolutionary stage when the energy
densities of matter and radiation are of the same order
(matter/radiation equality). Using the WMAP data for the
$\Lambda$CDM model \cite{WMAP}, we obtain $H_{eq}\sim
10^{-56}M_{Pl}$. Because the field oscillations start when the
Hubble parameter becomes less than the mass of the particles $(H
\lesssim m)$, particles with masses greater or less than
%$10^{-55}M_{Pl}$
$H_{eq}$ start to oscillate at the RD stage or at the MD stage,
respectively.} (the consistency condition is $\psi_{in} \lesssim
M_{Pl}$). Similarly, we obtain for all masses $m_{\psi} <
10^{-56}M_{Pl}$ (then the oscillations started at the beginning of
the MD stage) the bound
%%%%
\be{6b} \psi_{in} \lesssim M_{Pl}. \ee
%%%%
Here, $\psi_{in}$ denotes the amplitude of the initial field
oscillations of $\psi$ above the minimum position of the effective
potential. Usually, it is assumed that $\psi_{in} \sim O
(M_{Pl})$, although it depends on the form of $U_{eff}$ and can be
considerably less than $M_{Pl}$.  Particles with masses $m_{\psi}
\sim 10^{-33}$eV $\sim 10^{-61}M_{Pl}$ are of special interest
because via $\Lambda_{eff} \sim m_{\psi}^{2}$
\cite{GZ1,GZ,GZ(CQG1998),GZ(PRD2)} they are related to the
recently observed value of the effective cosmological constant
(dark energy) $\Lambda_{eff} \sim 10^{-123}\Lambda_{Pl(4)} \sim
10^{-57} \mbox{cm}^{-2}$. These ultra-light particles have a
period of oscillations $t \sim 1/m_{\psi} \sim 10^{18}$sec which
is of order of the Universe age. Thus, for these particles a
splitting of the scale factor of the internal space into a
background component and gravexcitons makes no sense. A more
adequate interpretation of the scale factor dynamics would be in
terms of a slowly varying background in the sense of a
%Paul Steinhardt's
quintessence scenario \cite{WCOS}, \cite{SahStar}.

Another very strong restriction on light gravexcitons follows from
experiments on the time variation of the fine structure constant
$\alpha$. It is well known (see e.g.
%\cite{Carroll,IJMPD})
%\cite{Dicke,Bek1,Carroll,MSK,SBM,Bek2}
\cite{Dicke}) that the interaction between a scalar field $\varphi
\equiv \kappa_0 \psi$ and an electromagnetic field $F$ of the form
$f(\varphi ) F^{2}$ results in a variation of the fine structure
constant $\alpha$:
%%%%
\begin{equation}\label{1d} \left|\frac{\dot{\alpha}}{\alpha}\right|\; =
\; \left| \frac{\dot{f}}{f}\right| \,.
\end{equation}
%%%%
(The dot denotes differentiation with respect to time.) This
relation has its origin in the observation that a theory with
$L=f(\varphi ) F^{2}$ can be interpreted as a theory of an
electromagnetic field in a dielectric medium with  permitivity
$\epsilon_d = f(\varphi)$ and permeability $\mu =
f^{-1}(\varphi)$. On the other hand, it is equivalent to a field
theory in vacuum with $L=\tilde{F}^2$ (an analog of $\tilde{F}$ in
our equation \rf{12e}) and a variable electric charge $e =
f^{-1/2}(\varphi)e_0$. Thus, the fine structure constant is also a
dynamical function: $\alpha = e^2/(\hbar c) = e^2_0/(f\hbar c)$.

Let $\varphi_0$ denote the value of $\varphi$ for a stable
configuration at the present time, and $\varphi- \varphi_0
=\kappa_0\eta = \eta /\bar{M}_{Pl}<1$ (here, $\bar{M}_{Pl} :=
M_{Pl}/\sqrt{8\pi}$) be small fluctuations in its vicinity. Then,
the interaction term of the Lagrangian reads: $\gamma (\eta
/\bar{M}_{Pl}) F^2$, where $\gamma := \left. df/d\varphi
\right|_{\varphi_0}$. In our case we have $\varphi_0 =0$, and
$f(\varphi )$ and $\gamma$ are defined by Eqs. \rf{11e} and
\rf{1a} as:
%%%%
\be{dop1} f(\varphi ) = -\frac12 e^{-2
\sqrt{\frac{d_1}{(D_0-2)(D-2)}}\; \varphi} \quad \Longrightarrow
\quad \gamma =  \sqrt{\frac{d_1}{(D_0-2)(D-2)}}\, .\ee
%%%%
The experimental bounds on the time variation of $\alpha$ have
been considerably refined during the last years (see, e.g.,
\cite{Hannestad,IPV,Webb,Melnikov,Uzan} and references therein).
Different experiments give different bounds on
$|\dot{\alpha}/\alpha|$ (see Table II in \cite{Uzan}), from
$\lesssim 10^{-12}\mbox{yr}^{-1}$ (following from the data
analysis of the observed cosmic microwave background
\cite{Hannestad}) to $\lesssim 10^{-17}\mbox{yr}^{-1}$ (following
from the Oklo experiment \cite{DD}). Estimates on primordial
nucleosynthesis require $|\Delta \alpha/\alpha| \lesssim 10^{-4}$
at a redshift of order $z = 10^9 - 10^{10}$ \cite{KPW}, i.e.
$|\dot{\alpha}/\alpha| \lesssim 10^{-14}\mbox{yr}^{-1}$. The WMAP
data
%computer simulation
analysis \cite{MMRTAV} gives upper bounds on the variation of
$\alpha$ during the time from re-ionization/recombination ($z \sim
1100$)  until today: $|\Delta \alpha/\alpha| \lesssim 2\times
10^{-2} - 6\times 10^{-2}$, i.e. $|\dot \alpha/\alpha | \lesssim
2\times 10^{-12} - 6\times 10^{-12}\mbox{yr}^{-1}$. In all these
estimates $\dot{\alpha} = \Delta \alpha /\Delta t$ is the average
rate of change of $\alpha$ during the time interval $\Delta t$
(corresponding to a redshift $z$). For our calculations we use the
estimate $|\dot{\alpha}/\alpha| \lesssim 10^{-15}\mbox{yr}^{-1}$
\cite{Webb} which follows from observations of the spectra of
quasars at a Hubble time scale $\Delta t \sim H^{-1} \sim 10^{10}$
years. For this bound, we obtain from \rf{1d}:
%%%%
\begin{equation}\label{2d}
\left| \frac{\dot{f}}{f}\right| = \left| \frac{1}{f}\frac{d f}{d
\varphi} \frac{\dot{\psi}}{\bar{M}_{Pl}} \right| = \left|
\frac{\gamma}{f} \frac{\Delta \psi}{\Delta t}
\frac{1}{\bar{M}_{Pl}}\right| = \left|\frac{\dot \alpha}{\alpha}
\right| \lesssim 10^{-15}\mbox{yr}^{-1} \; ,
\end{equation}
%%%%
which leads to the following restriction on the parameter
$\gamma$:
%%%%%
\begin{equation}\label{7} |\gamma|\approx \Delta t \left|
\frac{\dot{\alpha}}{\alpha}\right| \frac{\bar{M}_{Pl}}{\Delta
\psi} \; \Rightarrow \; |\gamma| \lesssim 10^{-5}.
\end{equation}
%%%
Here, we took into account the present value of $f(0) = 1$,
supposed for the time interval a value of $\Delta t \sim H^{-1}
\sim 10^{10}$ years, and assumed $\Delta \psi \sim \bar{M}_{Pl}$.
However, equation \rf{dop1} shows that in our model we have
$\left. \gamma \right|_{D_0=4} = \sqrt{d_1/[2(2+d_1)]} \sim
\mathcal{O}(1)$. This obvious contradiction means that light
gravexcitons with masses $m_{\psi} \lesssim 10^{-2}$GeV should
have decayed at sufficiently early times of the evolution of the
Universe in order  not to contradict the experimental bounds on
the variation of the fine structure constant\footnote{In general,
the performed estimates can be refined by accounting for
 viralization and accretion processes during the formation of overdense
 spacetime regions, such as galaxies \cite{BM1}. In this case, if the
 gravexcitons are also sufficiently clustered, the study of
 time variations of  $\alpha$ should be supplemented by an analysis of
 spatial variations of $\alpha$.}. From this
point of view, the presence of such light gravexcitons is
unacceptable for the time after the end of primordial
nucleosynthesis. Additionally, ultra-light gravexcitons can lead
to the appearance of a fifth force with characteristic length
scale $\lambda \sim 1/m_{\psi}$. Recent gravitational
(Cavendish-type) experiments (see e.g.
\cite{experiment2,experiment3}) exclude fifth force particles with
masses $m_{\psi}\lesssim 1/(10^{-2}\mbox{\rm cm})\sim 10^{-3}$eV.
This sets an additional restriction on the allowed mass region of
gravexcitons.

\subsection{Heavy gravexcitons\label{heavy}}

Let us now consider  gravexcitons with masses $m_{\psi} \gtrsim
10^{-2}$GeV. For such gravexcitons the decay \rf{3a}  plays an
important role. If there is no broad parametric resonance
("preheating") (see \cite{DL,Star1984,KLS1994} for the details of
the derivation), then for $H \ll m_{\psi}$ as result of this decay
the factor $e^{-\Gamma t}$ becomes dominant in the energy density
and the number density
%%%%
\be{13a} \rho_{\psi} \sim e^{-\Gamma t}a^{-3}\ , \qquad n_{\psi}
\sim e^{-\Gamma t}a^{-3}, \ee
%%%%%
and due to the gravexciton decay the Universe undergoes a
reheating up to the temperature \cite{GZ(PRD2),ENQ,CCQR}
\be{14}
T_{RH} \sim \sqrt{\frac{m^3_{\psi}}{M_{Pl}}}\, .
\ee
For a successful nucleosynthesis a temperature $T_{RH} \gtrsim
1$MeV is needed so that the gravexciton masses should satisfy
$m_{\psi} \gtrsim 10^4$GeV. This result is obtained under the only
assumption that gravexcitons dominate at the time of their decay.
%%%%%%%%%%%%
%%%%%%%%%%%%%%
Moreover, for a successful hot baryogenesis\footnote{A
low-temperature mechanisms for the baryogenesis can considerably
lower this mass value  \cite{ENQ,CKN} (see also \cite{MYY}).} it
should hold $m_{\psi} \gtrsim 10^{14}$GeV \cite{ENQ}. Thus, either
the decaying particles should have masses $m_{\psi }$ which
satisfy these lower bounds (and will decay before nucleosynthesis
due to the interaction channel shown in Fig. 1) or, for lighter
gravexcitons, we should find other scenarios which would allow us
to get rid off such particles before nucleosynthesis starts. The
latter can be achieved if the decay rate becomes larger. However,
a fast decay in the regime of a broad parametric resonance
\cite{KLS1994} may not be realized for gravexcitons since it
requires the condition $\Gamma >> H$ at the beginning of the decay
($H \sim m_{\psi}$) which is not satisfied (see estimate \rf{3a}).

Thus, in order to avoid the aforementioned problems, gravexcitons
should be sufficiently heavy and decay before nucleosynthesis
starts.
%There is another serious problem associated with heavy
%gravexcitons.
However, as was pointed out in Section 2, in models with
stabilized internal spaces the minimum of the effective potential
plays the role of the effective cosmological constant which
usually (for simplified cosmological models) is connected with the
gravexciton masses by a relation $\Lambda_{eff} \sim m_{\psi}^2$.
Hence, heavy gravexcitons result in $\Lambda_{eff} \gg
10^{-57}\mbox{cm}^{-2}$, what is in obvious contradiction to
recent observations \cite{Peebles}. For such particles a mechanism
should be found which could provide a reduction of $\Lambda_{eff}$
to its observable value. A similar problem with a large
cosmological constant exists also in superstring modular cosmology
\cite{BBS}. (For a discussion of the cosmological constant problem
(CCP) within the framework of string theory we refer to
\cite{W2}.) A possible resolution of this problem could consist in
a consideration of more or less realistic models which contain
different types of matter fields. Some of these fields should
violate the null energy condition (NEC) and the weak energy
condition (WEC).
%The
%relation $\Lambda_{eff} \sim m_{\psi}^2$ usually holds for
%simplified models.
As shown in our recent paper \cite{GMZ2}, in this case one can
obtain large gravexciton masses $m_{\psi} \sim
M_{*(4+D^{\prime})}$ and a small positive $\Lambda_{eff}$ which is
in agreement with the observed value\footnote{For a possible
resolution of the CCP in codimension-two brane-world scenarios we
refer to the recent work \cite{CCP}. }.

Most probably, primordial cosmological gravexcitons decayed
already at the early stages of the evolution of the Universe.
Nevertheless, it is of great interest to consider different
mechanisms which could lead to a gravexciton production   at
present time. Obviously, due to the Planck scale suppression, it
is hardly possible to observe the interaction  between
gravexcitons and photons in laboratory-scale experiments. Thus, we
should look for regions in our Universe where conditions could
exist which are suitable enough for such reactions to occur. For
example, a thermal gravexciton production would become possible at
temperatures $T \gtrsim m_{\psi}$ when the exponential suppression
in the creation probability is switched off: $P \sim \exp
(-m_{\psi}/T) \sim \mathcal{O}(1)$. Then, the creation rate (per
unit time per unit volume) can be estimated as $\nu \sim
T^6/M_{Pl}^2$ (see e.g. Ref. \cite{Rubakov} where a similar
estimate was given for the production of KK gravitons).
%Thus, number of
%particles created in volume $V$ (per unit time) is $\tilde \nu
%\times V$.
In our case, a thermal production of heavy gravexcitons with
masses $m_{\psi} \gtrsim 10^4$GeV would take place if $T >
10^{17}K$. The maximal temperatures which can be reached in any
known astrophysical objects\footnote{Concerning the maximal
possible temperature in the Universe, see the paper
\cite{Sakharov} by A. Sakharov where this temperature was
estimated as $T_{max} \sim T_{Pl} \sim 10^{32}$K.} (supernovas,
neutron stars, pulsars) are defined by nuclear reactions and do
not exceed $10^{10}-10^{12}\mbox{K} \sim 1-10^2$MeV. Hence, a
thermal gravexciton production in the cores of these objects is
absent. Such a production could be possible at early evolution
stages of the hot Universe, but the corresponding cosmological
gravexcitons would have decayed before nucleosynthesis started.
%(where
%$V \sim a^3$ is defined by the scale factor $a$ of the Universe).
\subsection{Gravexcitons and UHECR}

The observation of cosmic rays with ultra high energies (UHECR)
$\, E\gtrsim 10^{20}$eV (see e.g. review \cite{NW}) shows that our
Universe contains astrophysical objects where particles with
energies $E \gg m_{\psi} \gtrsim 10^4$GeV can be produced.
Obviously, these energies should have a non-thermal origin and
should be large enough for the creation of heavy gravexciton, e.g.
in reactions $2\gamma \rightarrow \psi$ ($\gamma-$quanta
conformally excite internal spaces). Then, if a gravexciton
acquires the energy $E > m_{\psi}$, it propagates through the
Universe without decay a distance \cite{GRS}:
\be{15}
l_D \sim \frac{E}{\Gamma \, m_{\psi}} \sim
\frac{E}{10^{20}{\mbox{eV}}}\left(
\frac{10^{17}{\mbox{GeV}}}{m_{\psi}}\right)^4l_{Pl}\, ,
\ee
where $\Gamma \sim m_{\psi}^3/M_{Pl}^2$ is defined by eq. \rf{3a}.
Substituting $E\sim 10^{20}$eV and $m_{\psi}\sim 10^4$GeV, we
obtain $l_D \sim 10pc$ which is considerably less than the
Greisen-Zatsepin-Kuzmin (GZK) cut-off distance $r_{GZK}\sim
50$Mpc. Thus, such heavy gravexcitons cannot be used for the
explanation of the UHECR problem. However, it can be easily seen
that for $m_{\psi}\lesssim 10^2$ GeV the decay length exceeds the
cut-off distance: $l_D \gg r_{GZK}$. On the other hand, the very
weak coupling constant $\kappa_0 \sim M_{Pl}^{-1}$ leads to a
negligible scattering of gravexcitons at the CMBR \cite{GRS}.
Thus, if a mechanism could be found to reconcile the presence of
gravexcitons of mass $m_{\psi}\sim 10^2$ GeV with a successful
nucleosynthesis, then such gravexciton could  be helpful in
solving  the UHECR problem.

\subsection{Gravexcitons and magnetars}

\begin{figure}[tbh]
\begin{center}
\epsfig{file=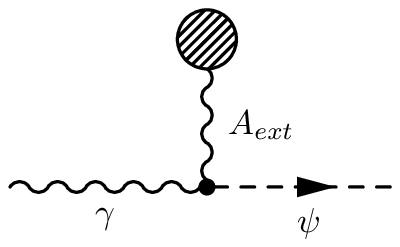,angle=0, width=4.5cm}
\end{center}
\caption{\label{fig2}Photon induced gravexciton creation in a
strong magnetic background field.}
\end{figure}

Another possible source for the production of gravexcitons could
be the strong magnetic fields of some astrophysical objects (e.g.
neutron stars, pulsars or magnetars). In analogy with effects
described for axions \cite{27,raff}, a strong external
electric/magnetic field can lead to oscillations between
gravitational excitons and photons in accordance with the diagram
of  Fig. \ref{fig2}.
%\vspace{-2.5cm}
\noindent It corresponds to an interaction term $L_{eff} \sim
\kappa_0 \psi F^{\mu \nu}_{ext} f_{\mu \nu}$, where $F^{\mu
\nu}_{ext}$ is the field strength of the external
electric/magnetic field and $f_{\mu \nu}$ denotes the photon wave
function. In the presence of a strong external field, the
probability for an electric/magnetic conversion of gravitational
excitons into photons (and vice versa) could be much greater then
\rf{3a} and would result in observable lines in the spectra of
astrophysical objects. Usually, the electric fields of
astrophysical objects are very small. However, the magnetic fields
can be very strong. For example, magnetars as special types of
pulsars can possess  magnetic fields of strengths
$B>B_{critical}\sim 4.4 \times 10^{13}$Gauss $\sim 3 \times
10^{-6}\mbox{GeV}^2$. So, it is of interest to estimate the
magnetic field strengths which would provide an efficient and
copious  gravexciton production. For this purpose we note that for
a creation of a gravexciton the energy conservation law in a
stationary external field requires an energy of the incident
photon of $E\gtrsim m_{\psi}$. Furthermore, an efficient
gravexciton production takes place for magnetic field strengths
$B^2 \sim m_{\psi}^2 \kappa_0^{-2} \sim m_{\psi}^2M_{Pl}^2$.
%where the factor $M_{PL}^2$ comes from the
%strength of the gravexciton-photon interaction.
Thus, for a gravexciton of mass $m_{\psi} \sim 10^4$GeV we obtain
$B \sim 10^{23}\mbox{GeV}^2$ which is much larger than
$B_{critical}$. Hence, we have to conclude that the magnetic field
strengths of magnetars are not sufficient for an efficient
production of gravexcitons. The reason for this is clear. First,
gravexcitons are heavy particles and, second, the
gravexciton-photon interaction itself is Planck-scale suppressed.

To conclude this section, we note that similar to the results on
dilatonic couplings $\sim e^{-\phi} F^2$ of electromagnetic fields
in inflation models \cite{21c} and string cosmology \cite{21b}, we
can expect in theory \rf{11e} an amplification of electromagnetic
vacuum fluctuations due to the presence of a dynamical gravexciton
background. Such amplification processes can generate sufficiently
strong magnetic seed fields to start galactic dynamo effects which
maintain the inter-galactic magnetic field at present time.
Furthermore, they can result in contributions to the observable
anisotropy of the cosmic microwave background.
%%%%%%%%%%%%%%%%%%%%%%%%%%%%%%%%%%%%%%%%%%%%%%%%%%%%%%%%%%%%%%%%%%%%%

\section{Conclusion\label{conclusion}}

In the present paper we investigated interactions of small
conformal scale factor fluctuations (gravitational excitons) of
extra-dimensional space components with 4D Abelian gauge fields.
The considered model was based on a factorizable background
geometry.

With the help of a toy model ansatz we demonstrated that the 4D
gauge invariant electromagnetic sector of the theory is
dilatonically coupled to the scale factor fluctuations
(gravexcitons). For a static background with background scale
factors frozen in the minimum of an effective potential, the
interaction term of gravexcitons and 4D-photons has the form
$(\psi/M_{Pl})F^{\mu \nu } F_{\mu \nu}$. This Planck scale
suppressed coupling leads for gravexcitons with mass $m_{\psi}$ to
a decay rate $\Gamma \sim m^3_{\psi}/M^2_{Pl}$. Accordingly,
gravexcitons are WIMPs with respect to this interaction channel.
Depending on the concrete gravexciton mass  different physical
effects will become relevant.
\begin{itemize}
\item
Light gravexcitons with masses $m_{\psi} < 10^{-2}$ GeV have a
life time which exceeds the age of the Universe so that they can
be considered as Dark Matter (DM). In this case, the inequality
\rf{6a} is saturated and in order that gravexcitons constitute the
present DM, each mass value $m_{\psi}$ should be connected with a
corresponding special value of the initial fluctuation amplitude
$\psi_{in}$. Additionally, laboratory tests of the gravitational
inverse-square law limit the gravexciton mass from below:
$m>10^{-3}$eV. Furthermore, such gravexcitons lead to a temporal
variability of the fine structure constant above the
experimentally established value (even with the Webb et al. data
\cite{Webb} taken as upper limits on the variability of $\alpha$).
Thus, this case seems to be excluded\footnote{Apart from the
recent radical hypothesis that $\alpha$ is significantly spatially
dependent so that its measured value is characteristic for our
spatial vicinity only \cite{BM1}.}.
\item
Ultra-light gravexcitons with masses $m_{\psi} \sim 10^{-33}$ eV
are closely related to the observable cosmological constant/dark
energy: $m_{\psi}^2 \sim \Lambda_{obs}\sim
10^{-57}\mbox{cm}^{-2}$. The oscillation period of these particles
exceeds the age of the Universe and a splitting of the scale
factor of the internal space into a background component and
gravexcitons makes no sense. A more adequate interpretation of the
scale factor dynamics would be in terms of a slowly varying
background similar to the scalar field dynamics in a quintessence
scenario. However, any dark energy model based on ultra-light
gravexcitons should first address the problem of absence of a
"fifth force".
\item
For gravexcitons with masses $m_{\psi} > 10^{-2}$ GeV, decay
processes play an important role during the evolution of the
Universe. These processes can considerably contribute to
reheating. We demonstrated that heavy gravexcitons with masses
$m_{\psi} \gtrsim 10^4$GeV can best meet the existing cosmological
restrictions. On the one hand, the reheating temperature of these
particles is sufficiently high for a successful nucleosynthesis.
On the other hand, they intensively decay already before
nucleosynthesis starts, what prevents a too large variation of the
fine structure constant afterwards.
\end{itemize}
The Planck-scale suppression of the interaction and the high
energy  $E\gtrsim m_{\psi}$ needed for the creation of heavy
gravexcitons with masses $m_{\psi} \gtrsim 10^4$GeV will  make it
very difficult, or even impossible, to observe such reactions
 in future laboratory-scale experiments.  So, it is of interest to
 clarify whether  reactions
with intensive gravexciton production may take place in some
astrophysical objects.

As first example we discussed a possible contribution of
gravexcitons to UHECR with energies $E\sim 10^{20}$eV. It is clear
that, due to their weak interaction with ordinary matter,
gravexcitons will propagate in the Universe without significant
scattering on CMBR. This makes them attractive as a possible
candidate for UHECR. Our estimates show, that the decay length of
gravexcitons with masses $m_{\psi} \gtrsim 10^4$ GeV is very
short, but for masses $m_{\psi} \lesssim 10^2$ GeV it becomes much
longer than the Greisen-Zatsepin-Kuzmin cut-off distance.

As second example we considered the interaction between
gravexcitons and photons in the strong magnetic background fields
of  magnetars. The presence of the background fields will strongly
enhance the reaction and, in analogy to axion-photon oscillations,
one might expect the occurrence of gravexciton-photon
oscillations. However, our estimates show that even the strong
magnetic fields of magnetars with field strengths
$B>B_{critical}\sim 4.4 \times 10^{13}$ Gauss are not strong
enough for an efficient gravexciton production.

It remains to clarify whether other astrophysical objects or other
interaction channels could lead to an efficient gravexciton
production with directly observable phenomena.

\vspace{1ex}\mbox{}\\ {\bf Note added} After finishing this paper,
we noticed that the paper \cite{kolb2} appeared where cosmological
implications of extra dimensions are also discussed.

%%%%%%%%%%%%%%%%%%%%%%%%%%%%%%%%%%%%%%%%%%%%%%%%%%%%%%%%%%%%%%%%%%%%%
\vspace*{1ex}

\mbox{} \\ {\bf Acknowledgments}\\ We thank V.A. Rubakov, Yu.N.
Gnedin  and G.S. Bisnovaty-Kogan for useful discussions during the
preparation of this paper. U.G. acknowledges support from DFG
grant KON/1344/2001/GU/522. A.S. was partially supported by the
Russian Foundation for Basic research, grant 02-02-16817, and by
the Research program "Elementary particle physics and fundamental
nuclear physics" of the Russian Academy of Sciences. The work of
A.Z. was partly supported by the programme SCOPES (Scientific
co-operation between Eastern Europe and Switzerland) of the Swiss
National Science Foundation, project No. 7SUPJ062239.
%%%%%%%%%%%%%%%%%%%%%%%%%%%%%%%%%%%%%%%%%%%%%%%%%%%%%%%%%%%%%%%%%%

%%%%%%%%%%%%%%%%%%%%%%%%%%%%%%%%%%%%%%%%%%%%%%%%%%%%%%%%%%%%%%%%%%%

%%%%%%%%%%%%%%%%%%%%%%%%%%%%%%%%%%%%%%%%%%%%%%%%%%%%%%%%%%%%%%%%%%

\end{document}